# Spin-orbit-free Weyl-loop and Weyl-point semimetals in a stable three-dimensional carbon allotrope

Yuanping Chen[1,2*], Yuee Xie[1], Shengyuan A. Yang[3†], Hui Pan[4],

Fan Zhang[5‡], Marvin L. Cohen[6] and Shengbai Zhang[2]

[1]*Department of Physics, Xiangtan University, Xiangtan, Hunan 411105, China*
[2]*Department of Physics, Applied Physics, and Astronomy, Rensselaer Polytechnic Institute, Troy, New York 12180, USA*
[3]*Research Laboratory for Quantum Materials and EPD Pillar, Singapore University of Technology and Design, Singapore 487372, Singapore*
[4]*Department of Physics, Beihang University, Beijing 100191, China*
[5]*Department of Physics, University of Texas at Dallas, Richardson, Texas 75080, USA*
[6]*Department of Physics, University of California at Berkeley, and Materials Sciences Division, Lawrence Berkeley National Laboratory, Berkeley, California 94720, USA*

## Abstract

Topological band theory has revolutionized our understanding of electronic structure of materials, in particular, a novel state - Weyl semimetal - has been predicted for systems with strong spin-orbit coupling (SOC). Here, a new class of Weyl semimetals, solely made of light elements with negligible SOC, is proposed. Our first-principles calculations show that conjugated *p* orbital interactions in a three-dimensional pure carbon network, termed interpenetrated graphene network, is sufficient to produce the same Weyl physics. This carbon allotrope has an exceptionally good structural stability. Its Fermi surface consists of two symmetry-protected Weyl loops with linear dispersion along perpendicular directions. Upon the breaking of inversion symmetry, each Weyl loop is reduced to a pair of Weyl points. The surface band of the network is nearly flat with a very large density of states at the Fermi level. It is reduced to Fermi arcs upon the symmetry breaking, as expected.

The discovery of topological insulators has revolutionized our understanding of fundamental properties of crystalline materials, in which nontrivial topology encoded in the band structure gives rise to many intriguing physical effects [1,2]. More recently, it was recognized that the topological classification could be generalized beyond insulators, leading to the notion of topological semimetals [3-18]. In particular, a Weyl semimetal only has pairs of doubly degenerate points at the Fermi energy, known as Weyl points, around which the low-energy quasiparticles can be described by a $2 \times 2$ Weyl Hamiltonian [3-13]. This is distinct from those semimetals or metals where both Weyl points and conventional bands coexist at the Fermi energy [16-18]. Each Weyl point is a monopole in momentum space and enjoys a topological protection: integrating the Berry flux over a constant energy surface enclosing the point yields a quantized Chern number corresponding to the topological charge of the Weyl point. Weyl semimetals are anticipated to reveal remarkable physical properties, e.g., linear band dispersions, high carrier mobility, protected surface Fermi arcs [5], and chiral and axial anomalies [19,20] long-speculated in high-energy physics. The search for Weyl semimetals in solid state systems has been intense. Proposals so far are exclusively focused on those compounds involving heavy elements, with the motivation that a strong spin-orbit coupling (SOC) could enable the desired nontrivial band structure.

On the other hand, carbon materials, ranging from fullerenes, to carbon nanotubes, and to graphene, have been a focus of materials science and condensed matter physics for decades [21-24]. The success of monolayer graphene has stimulated the interest in searching for various carbon allotropes [25-34], which use graphene as basic building blocks. Besides the carbon nanotube (as a rollup of graphene) and graphite (as a stack of graphene), graphene can also form three-dimensional (3D) carbon structures through interpenetration of layers [30-34], thereby exhibiting an ordered alternation of $sp^2$- and $sp^3$-hybridization. What has been lacking in such a search is, however, a deeper understanding of the origin for the fascinating functionalities. Although it is well known that the uniqueness of graphene lies in its gapless Dirac cone, as a result of conjugated $p_z$ orbitals, the effect of 3D conjugation of $p$ orbitals, now including $p_x$, $p_y$, and $p_z$, has not been recognized. In essence, the gapless Dirac cone is a two-dimensional (2D) rendering of the more general 3D Weyl cone. Hence, it is natural to ask whether a 3D conjugation of the $p$ orbitals will lead to the parent, so far hidden, Weyl semimetals and related structures. As this is a pure orbital effect, the Weyl semimetals here would be a new class of spin-orbit-free topological matter. Note that our discussion is not limited to pure carbon, nor just to the $p$ orbitals. In other words, the effect can as well been observed in organic frameworks [35,36] with conjugation or even for half-occupied transition metal $d$ orbitals.

In this Letter, we report the prediction of such a new matter, based on first-principles calculations of a 3D carbon crystalline allotrope. The allotrope has an exceptionally good structural stability and its Fermi surface is made of two Weyl loops [7,12-15], with linear band dispersions along perpendicular directions. For a finite-sized sample, a nearly-flat band may form on the surface occupying the entire region in the surface Brillouin zone (BZ) enclosed by the surface-projected Weyl loops. This leads to a huge surface density of states (DOS) at the Fermi level. Each Weyl loop is protected by a chiral symmetry and a topological winding number [13]. When purposely breaking the inversion and chiral symmetries, each Weyl loop is transformed into a pair of Weyl points, along with a Fermi arc when a surface is formed. Both the shape of the Weyl loops and the positions of the Weyl points are sensitive to uniaxial strain; a pair of them with opposite chirality can even mutual annihilate at a single point.

Our proposed 3D carbon consists of strongly interlocked graphene sheets which share zigzag atomic chains extending along the *c*-axis, as shown in Fig. 1. The structural symmetry belongs to the space group *CMCM* ($D_{2h}$-17), and the Bravais lattice is a simple monoclinic system with an optimized angle $\theta \cong 86.6°$ between the two lattice vectors in the basal plane. Each primitive unit cell contains six C atoms, as shown in the inset of Fig. 1, which form two separate obtuse triangles symmetrically placed with respect to an inversion center at the center of the unit cell. The six atoms can be further divided into two groups: the two near the center of the cell which are fourfold-coordinated (marked in grey in the inset of Fig. 1); and the other four which are threefold-coordinated (marked in red). Therefore, the bonding character of this material is somewhere between that of diamond with $sp^3$ hybridization and that of graphene (or graphite) with $sp^2$ hybridization.

Our first-principles calculations were based on the density functional theory [37-40] (see the Supplemental Material). The structurally optimized lattice constants are $a = b = 4.33$ Å and $c = 2.47$ Å, respectively. The length of the bonds (1.55 Å) connecting the fourfold coordinated carbon atoms is close to that of diamond, whereas that (1.43 Å) connecting the threefold coordinated carbon atoms are close to that of graphene. The bond angles vary from 106 to 111°, comparable to 109.47° for diamond. Although the structure appears quite porous, it exhibits an exceptional stability. The calculated cohesive energy is $E_{\text{coh}} = 7.62$ eV/C. Although it is slightly smaller than the values for graphite (7.85 eV/C) and diamond (7.72 eV/C), it is larger than the values of C60 (7.48 eV/C) and most other 3D carbon allotropes, such as bct-C4 (7.54 eV/C) [41], carbon Kagome lattice (7.44 eV/C) [34], and M-carbon (7.20 eV/C) [42]. The large $E_{\text{coh}}$ and bulk modulus of 311 GPa indicate that the structure is rather stable. The calculated phonon spectrum, shown in Fig. S1 of the Supplemental Material, reveals the absence of any imaginary frequencies over the entire BZ, suggesting that the structure is also dynamically stable.

In the electronic band structure shown in Fig. 2(a), the linear dispersions along the Γ-Z and Y-T through the high-symmetry points are observed near the Fermi level. A closer look reveals that the linear band-crossing in fact occurs along two closed loops traversing the BZ, as schematically shown in the upper left panel of Fig. 2(c). The two loops reside in the (110) mirror invariant plane around $k_c = \pm 0.45\pi/c$ and are time-reversal or inversion images of each other. To better illustrate this, we show in Fig. 2(a) the band dispersion between T' and Y', with the T'-Y' line bisecting the Z-T and Γ-Y lines. Here, the linear band-crossing along T'-Y' can also be clearly seen. The linear dispersion is expected to give rise to a vanishingly small DOS around the Fermi level, as indeed observed in the right panel of Fig. 2(a), in which the partial density of states (PDOS) projected to different atomic orbitals is shown. One notices that the bands around the Fermi level are mainly made of the $p_x$ and $p_y$ orbitals, spatially located on only one type of atom, i.e., the four peripheral carbon atoms. In the inset of Fig. 2(a), a charge analysis further reveals that the two orbitals are made of π-bonds with $sp^2$ hybridization. The structure, the linear dispersion, and the π-band character are all reminiscent of 2D graphene. If one ignores the small variation in $k_c$ on the band-crossing line, the energy spectrum may be viewed as a superposition of staked 2D graphene along the [110] direction.

In the above analysis, we did not consider electron spin because for carbon the SOC is known to be negligibly small. As such, the two decoupled spin sectors can be treated as two identical copies, leading to a trivial twofold degeneracy for the energy bands. In this sense, although each band-crossing loop is fourfold degenerate, it is in fact composed of two identical loops with twofold degeneracy, and hence should be termed as a Weyl loop (or Weyl ring).

To capture the band-edge features found in the first-principles calculations, we construct a minimal tight-binding model that includes the four peripheral (red colored) carbon atoms with one $p$ orbital per atom. Taking into account the nearest-neighbor and second-neighbor hopping processes, we arrive at the following Hamiltonian,

$$\mathcal{H}(\boldsymbol{k}) = \begin{bmatrix} 0 & Q(\boldsymbol{k}) \\ Q^\dagger(\boldsymbol{k}) & 0 \end{bmatrix}, \quad \text{with} \quad Q(\boldsymbol{k}) = \begin{bmatrix} f_{14} & f_{13} \\ f_{24} & f_{23} \end{bmatrix}, \qquad (1)$$

where $f_{ij}(\boldsymbol{k}) = \sum_\mu t_{ij} e^{-i\boldsymbol{k}\cdot\boldsymbol{d}_{ij}^\mu}$, $i,j \in \{1,2,3,4\}$ are the site labels in the inset of Fig. 1, $t_{ij}$ is the hopping strength between sites $i$ and $j$, $\boldsymbol{d}_{ij}^\mu$ is the vector directed from $j$ to $i$, and $\mu$ runs over all equivalent lattice sites under translation. The spectrum of energy band is symmetric about zero energy because of the presence of a chiral (sublattice) symmetry $\mathcal{C} = \sigma_z \otimes \sigma_0$ for Eq. (1) such that $\mathcal{C}H\mathcal{C}^{-1} = -H$ independent of $\boldsymbol{k}$, where $\sigma_\alpha$ are the Pauli matrices. It is easy to find that zero-energy states would appear if the following two conditions are satisfied: (1) $k_a = k_b$ and (2) $\cos(k_c c/2) = \sqrt{t_{13} t_{24}/(4 t_{14} t_{23})}$. The first condition restricts the zero-energy states to the mirror invariant plane, whereas the second one further restricts them onto two separate loops at $\pm K_c = (2/c)\arccos[\sqrt{t_{13} t_{24}/(4 t_{14} t_{23})}]$. Despite its simplicity, the model captures the essential physics of the first-principles results. The fitted band structure of the tight-binding model is shown in Fig. 2(a), and it agrees well with the first-principles results.

The linear dispersion around the Weyl loops can be explicitly demonstrated by deriving the low energy effective Hamiltonian. Around the loop at $\tau_z K_c$ ($\tau_z = \pm 1$), the effective Hamiltonian is

$$H_{\text{eff}}(k_\perp, k_c) = -v_\perp k_\perp \sigma_y + \tau_z v_c k_c \sigma_x, \qquad (2)$$

where the wave-vector $\boldsymbol{k} = (k_\perp, k_c)$, staying in the plane perpendicular to the loop, is measured from the band-crossing point, and $v_\perp = (at_{13})/2$ and $v_c = (c/2)\sqrt{4t_{14}^2 - t_{13}^2}$. For each loop, Eq. (2) takes the form of a 2D Weyl Hamiltonian similar to that of graphene, though having anisotropy, and the chirality reverses between the two loops as required by the time reversal symmetry. Also similar to graphene, since the band gap is inverted for the $k_c$ values in the two regions separated by the Weyl loops, there should be surface bands [13] in either one of the regions depending on the boundary condition. Figure 3 shows the first-principles results for our carbon system with a 10-layer thick slab geometry along the [010] direction. The band near the Fermi level, marked in red, is localized on the surface and spans the whole region in the 2D surface BZ between the two surface-projected Weyl loops, as indicated by the pink-colored region in the inset of Fig. 3. If both the crystalline bulk and the surface approximately respect the chiral symmetry and its momentum independence, the surface band dispersion would be almost flat and contribute to a huge surface DOS (see the right panel in Fig. 3). This could give rise to interesting electron interaction effects, such as magnetic ordering and 2D superconductivity with high $T_c$.

The protection of a Weyl loop can be inferred from the 1D winding number (i.e., quantized Berry phase in units of $\pi$) of $\mathcal{H}(\mathbf{k})$ along a close path $\mathcal{L}$ encircling the Weyl loop [13]: $N_\mathcal{L} = \frac{1}{2\pi i}\oint_\mathcal{L} dl \cdot \text{Tr}[Q^{-1}(\mathbf{k})\nabla_l Q(\mathbf{k})] = \pm 1$. Note that the quantization of Berry phase is possible only when the chiral symmetry $\mathcal{C}$ is present [43]. In general, higher-order hopping processes may produce some momentum dependence of this symmetry and strong SOC even violate it. As a consequence, the Weyl loops in the first-principles results in Fig. 3 deviate slightly from $E(\mathbf{k}) = 0$.

Weyl loops in 3D have also been found in photonic crystals [12], nodal superconductors [13], superlattice structures [7,14], and doped narrow-gap insulator [15] with more delicate chiral symmetries. Perturbations that preserve the chiral symmetry can only distort or displace the Weyl loops but cannot destroy them, although the Weyl loops with opposite winding numbers can pair-annihilate. In the present case, there is also an additional mirror symmetry, which confines the Weyl loops in the mirror invariant plane but is not essential for their topological protection.

As an example, consider applying a uniaxial strain along the $[1\bar{1}0]$ crystal direction. This changes the angle $\theta$ but preserves the chiral symmetry as well as the mirror symmetry. As a result, we find that the two Weyl loops still stay in the mirror invariant plane but move closer to the BZ center. As shown in Fig. 2(c), once $\theta$ is less than 69°, the two loops merge into a single loop around the BZ center, which shrinks as $\theta$ further decreases and eventually the spectrum is gapped out when $\theta$ goes below 64°. Note that the point just before the loop disappears (the white dot in the bottom right panel of Fig. 2(c)) is not a Weyl point, because the dispersions around it are quadratic in the mirror invariant plane.

Weyl points can only be obtained when inversion or time reversal symmetry is broken. Since carbon based materials are non-magnetic, we choose to break inversion symmetry to demonstrate the existence of a Weyl semimetal phase. In such a case, the total number of Weyl points must be a multiple of four. This is because for each Weyl point at $k$, there must be a partner Weyl point at $-k$ with the same chirality, as required by time reversal symmetry, meanwhile the total chirality must add up to zero, implying the existence of at least two other Weyl points with opposite chirality. We now consider two scenarios. In the first one, we artificially displace one of the four peripheral carbon atoms in a unit cell (see Fig. 4(a)). Consequently, four Weyl points indeed appear, as shown in Fig. 4(c). Since the mirror symmetry is also broken, these points are off the mirror invariant plane in general. In the second scenario, we insert He atoms into the holes of the carbon network with a filling of one atom per cell (Fig. 4(b)). Our structural optimization reveals that the He atom prefers to stay at a location slightly off the center of the hole, thereby breaking inversion symmetry. This also leads to four Weyl points, as seen in Fig. 4c. We can further move these Weyl points around in the BZ by, e.g., applying a uniaxial strain along either [100] or [010] direction, as indicated in Fig. 4(b). It is noted that the Weyl points here are distinct from those whose presence requires mirror and rotational symmetries [44].

Once a Weyl point is created, it would be topologically stable, as it is locally protected by the Chern number of a constant energy surface enclosing the Weyl point [3-13]. For an open system with a surface, surface Fermi arcs must exist connecting the surface-projected Weyl points of opposite chiralities. We calculate the energy spectrum for the second scenario with a slab geometry and indeed confirm the existence of Fermi arcs. Figure 4(d) shows schematically the Fermi arcs for the (100) surface. Again, we stress that in the above discussion, electron spin is a dummy degree of freedom because of the negligibly small SOC strength. So the fourfold degenerate point at Fermi energy are composed of two elementary Weyl points (one for each spin species) with the same chirality. Therefore, it should be regarded as a Weyl point with a Chern number "2=1+1" rather than being considered as a Dirac point [9,11,13].

In conclusion, based on first-principles calculations we propose a new family of matter to solid state obeying the spin-orbit-free Weyl physics, as well as adding a member to the family of carbon allotropes with novel topological properties and good stability. Remarkably, the Fermi surface of our carbon allotrope is a pair of Weyl loops, which leads to flat surface band with large surface DOS around Fermi energy. Each Weyl loop is chiral-symmetry protected by the quantized Berry phase. We also demonstrate that, upon breaking of inversion and chiral symmetries, each Weyl loop will split into a pair of Weyl points, connected by a Fermi arc when a surface is formed. Finally, we show that uniaxial strain can be used to tune the shape of the Weyl loops and the positions of the Weyl points to a degree that a pair of them with opposite chirality can mutual annihilate at a single point in the BZ.

We thank H. Wang, D. West and D. L. Deng for useful discussions. YC and YX acknowledge support by the National Natural Science Foundation of China (Nos.11474243, 51376005 and 11204262). SAY acknowledges support by funding SUTD-SRG-EPD2013062. FZ acknowledges support by UT Dallas research enhancement funds. MLC acknowledges support by NSF Grant No. DMR-10-1006184, and the theory program at the Lawrence Berkeley National Laboratory through the Office of Basic Science, US DOE under Contract No. DE-AC02-05CH11231. SZ acknowledges support by US DOE under Grant No.DE-SC0002623.

*Corresponding email:\*chenyp@xtu.edu.cn; †shengyuan_yang@sutd.edu.sg; ‡zhang@utdallas.edu.*

**Figure Captions**

Fig. 1 (color online). A perspective view of the carbon atomic structure. Inset: The unit cell consisting of two (grey) atoms with $sp^3$-hybrization and four (red) atoms with $sp^2$-hybrization (labeled 1 to 4). *a*, *b*, and *c* in the bottom right coordinate system are lattice parameters, and $\theta$ is the angle between *a* and *b*.

Fig. 2 (color online). Band structure and Weyl loops. (**a**) Band structure and orbital-projected PDOS. Black solid lines are the DFT results, whereas the orange dotted lines are the tight binding results with $t_{13} = t_{24} = -0.65$ eV and $t_{14} = t_{23} = 3$eV. Inset: Wavefunction for the state labeled by a red point. (**b**) The BZ of system in Fig. 1. (**c**) Weyl loops inside the (110) mirror invariant plane of the BZ for different angle $\theta$, which is changed by applying a strain along the $[1\bar{1}0]$ direction. The color bar indicates the energy difference between conduction band and valence band at each $k$ point.

Fig. 3 (color online). Band structure and DOS of a slab with 10 layers along the [010] direction. Left panel: The red line indicates the surface band near the Fermi level. Inset: A schematic plot showing the region of the 2D BZ (in pink) for the corresponding surface band. This region is enclosed by the surface-projected Weyl loops (in red). Right panel: DOS of the surface layer (red line) and a graphene layer in bulk (blue line).

Fig. 4 (color online). Weyl points obtained by breaking inversion symmetry. (**a**) Moving one atom in the unit cell. (**b**) Inserting He atoms into the holes of the carbon network. The He atoms prefer stay at locations slightly off the hole centers. (**c**) Schematic figure of the four resulting Weyl points in the BZ. (**d**) Schematic figure showing the surface Fermi arcs in the surface BZ connecting pairs of the projected Weyl points with opposite chiralities.

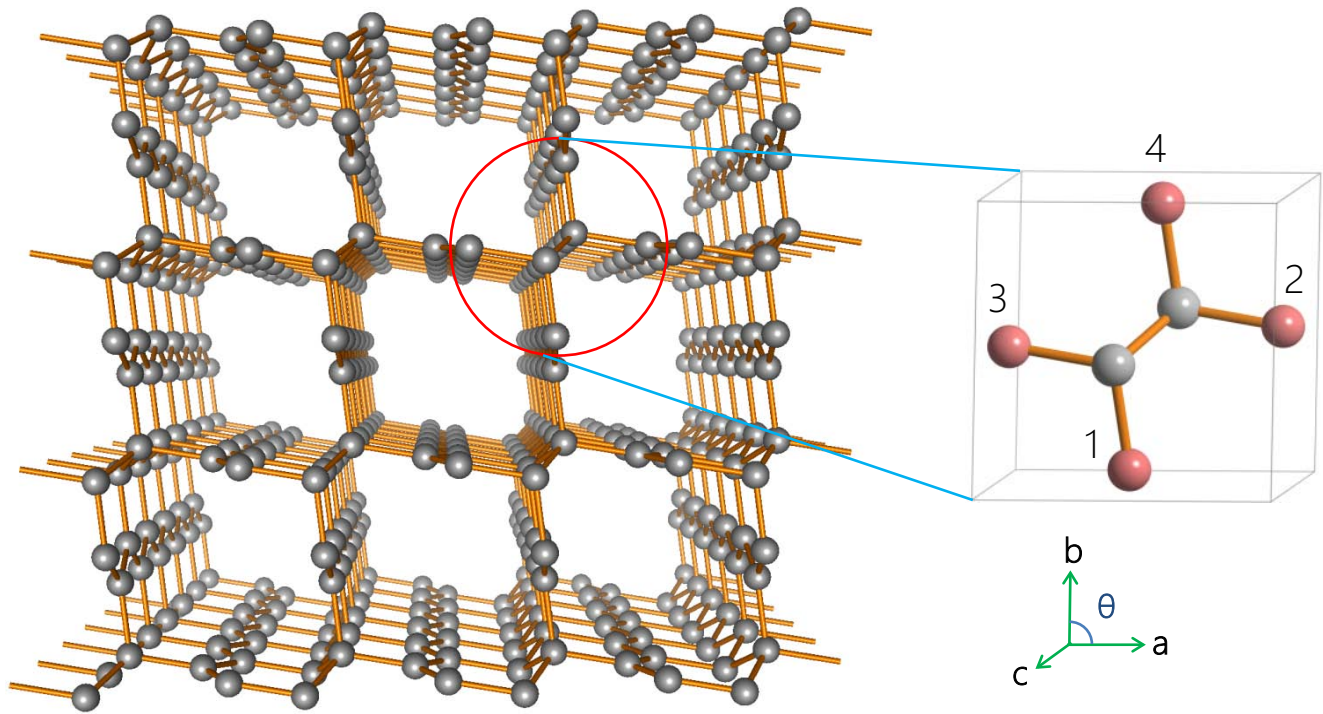

Figure 1

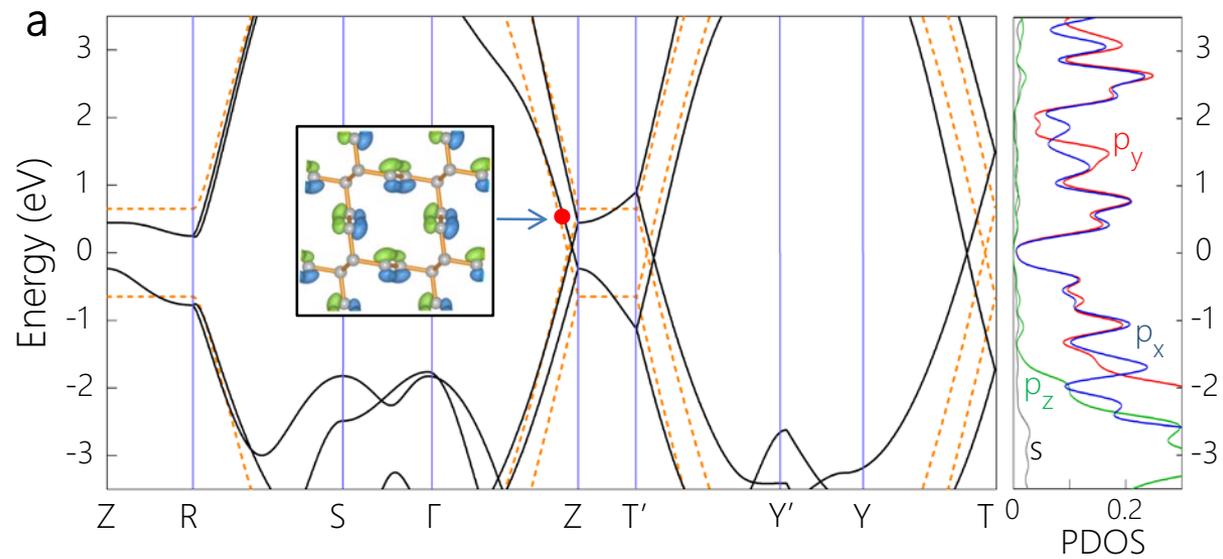
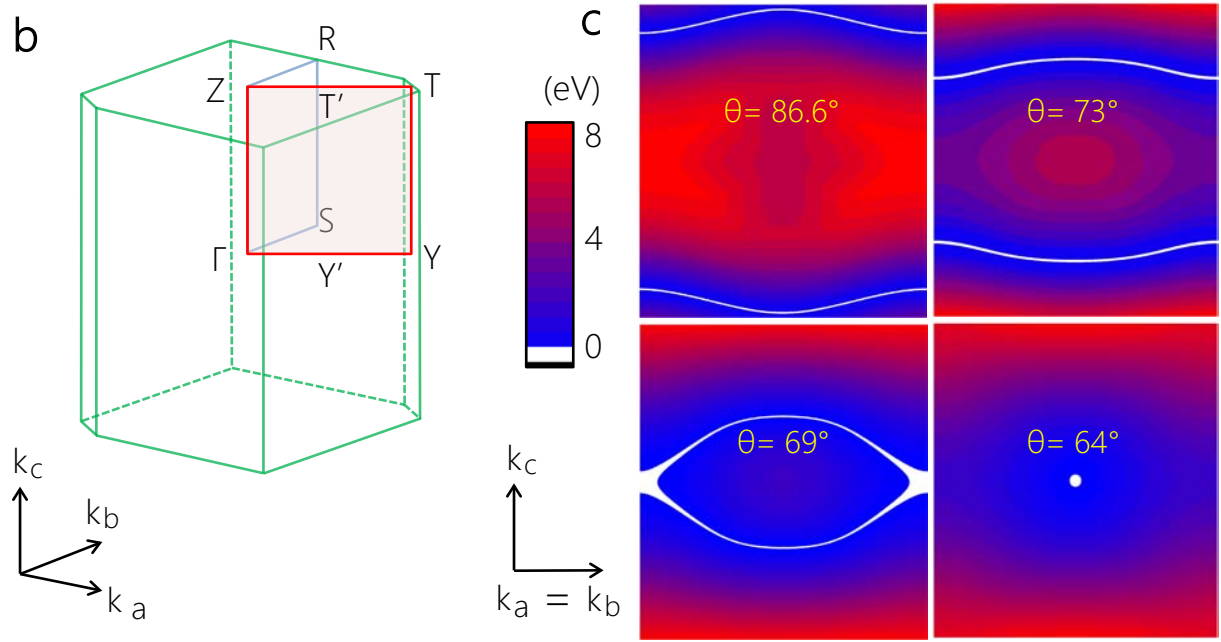

Figure 2

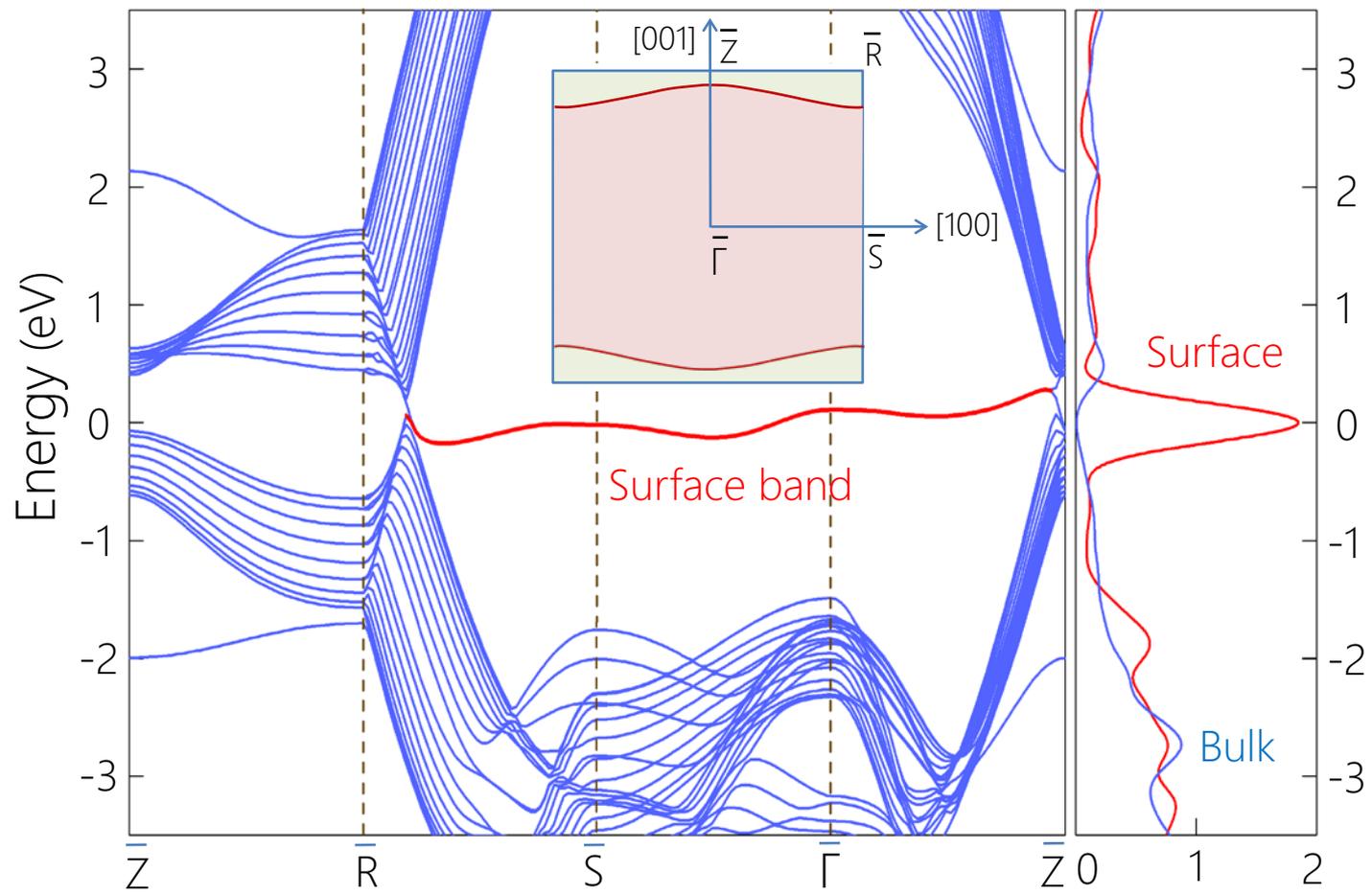

Figure 3

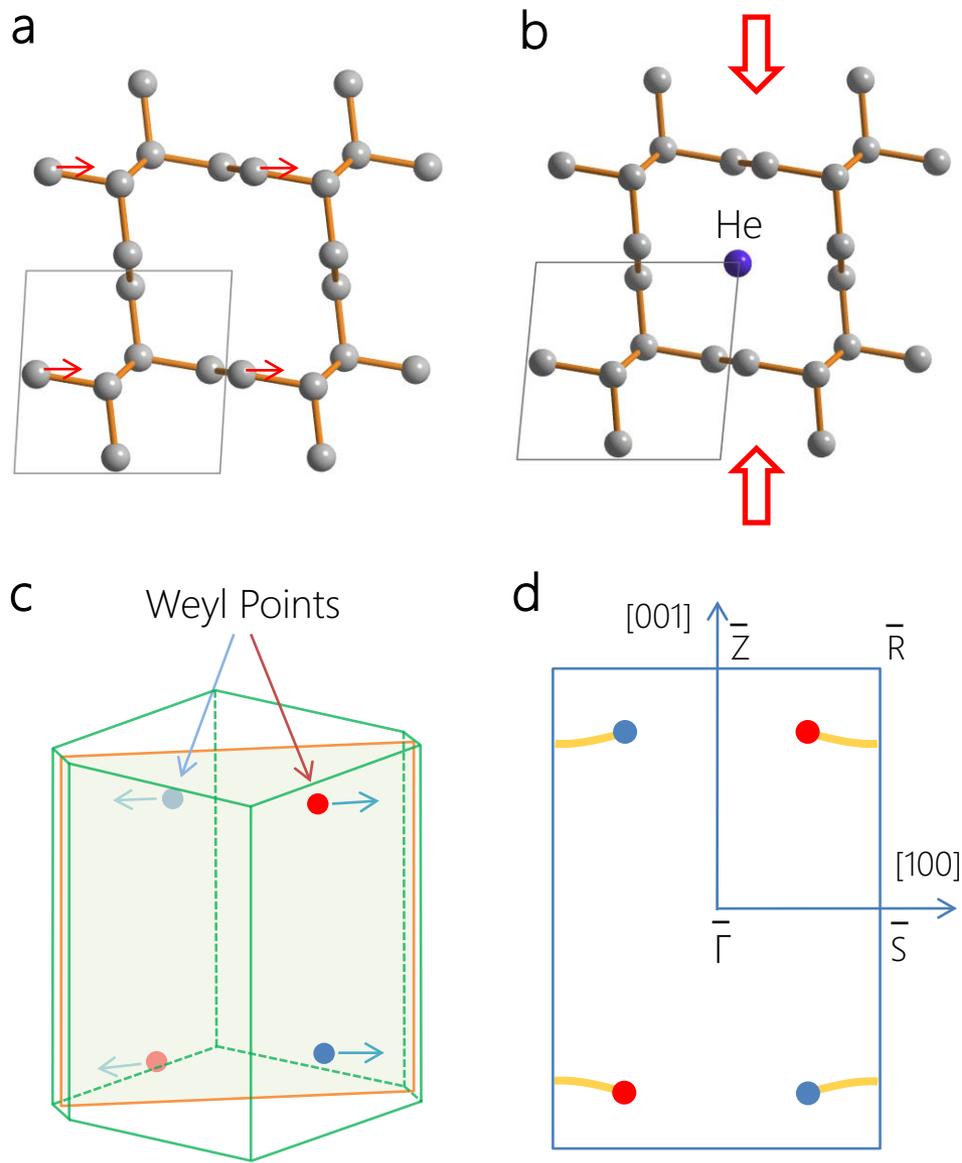

Figure 4

# Supplementary Information (SI)

**Methods**

Our first-principles calculations were based on the density functional theory (DFT) with the Perdew-Burke-Ernzerhof (PBE) approximation for the exchange-correlation functional [1]. The core-valence interactions were described by the projector augmented wave (PAW) potentials [2], as implemented in the VASP code [3]. Plane waves with a kinetic energy cutoff of 500 eV were used as the basis set. Slab calculations were carried out using periodic supercells [4]. All atoms were relaxed until the forces were smaller than 0.01 eV/Å. And a $9 \times 9 \times 16$ $k$-point mesh was used for the BZ integration. It is noted that in all calculations we turned off the spin-orbit couplings (SOC), because the effect of SOC in carbon structures is negligible. We also compare all the results with those in the presence of SOC, and find no noticeable differences between the two cases. This directly implies the negligibly weak SOC in our carbon structure.

**Supplementary figure**

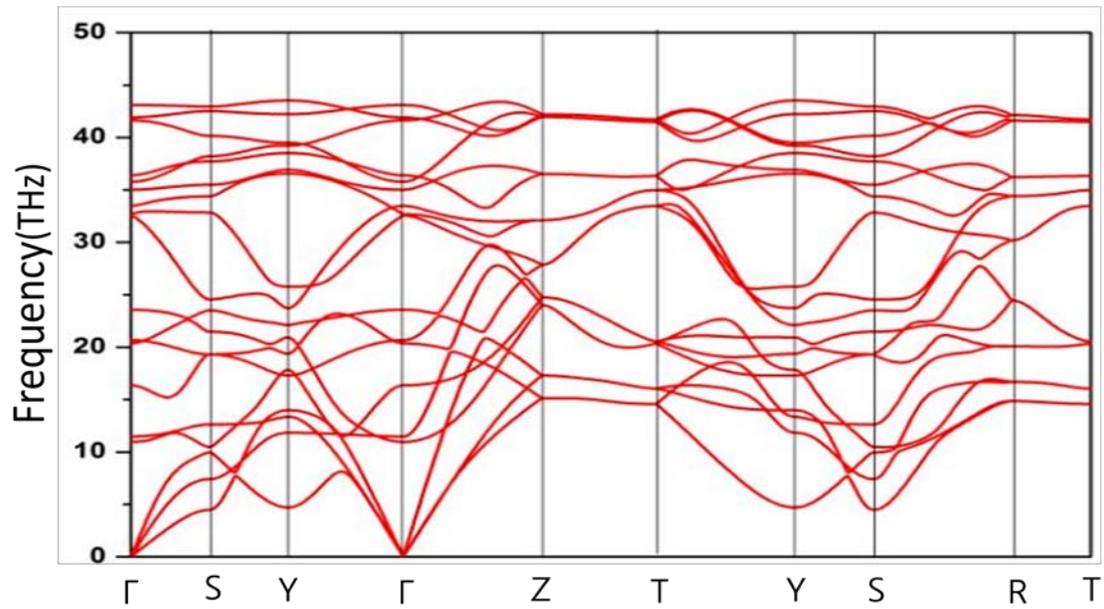

Figure S1 Phonon dispersion of the interpenetrated graphene network in Fig. 1. It reveals the absence of any imaginary frequencies over the entire BZ, suggesting that the structure is dynamically stable.